# Water Permeation through Layered Graphene-based Membranes: A Fully Atomistic Molecular Dynamics Investigation

Daiane D. Borges[1], Cristiano F. Woellner[1,2], Pedro A. S. Autreto[1], and Douglas S. Galvao[1]
[1]Applied Physics Department, University of Campinas - UNICAMP, Campinas-SP 13083-959, Brazil
[2]Department of Materials Science and Nano Engineering, Rice University, Houston, Texas, USA

## ABSTRACT

Graphene-based membranes have been investigated as promising candidates for water filtration and gas separation applications. Experimental evidences have shown that graphene oxide can be impermeable to liquids, vapors and gases, while allowing a fast permeation of water molecules. This phenomenon has been attributed to the formation of a network of nano capillaries that allow nearly frictionless water flow while blocking other molecules by steric hindrance effects. It is supposed that water molecules are transported through the percolated two-dimensional channels formed between graphene-based sheets. Although these channels allow fast water permeation in such materials, the flow rates are strongly dependent on how the membranes are fabricated. Also, some fundamental issues regarding the nanoscale mechanisms of water permeation are still not fully understood and their interpretation remains controversial. In this work, we have investigated the dynamics of water permeation through pristine graphene and graphene oxide model membranes. We have carried out fully atomistic classical molecular dynamics simulations of systems composed of multiple layered graphene-based sheets into contact with a water reservoir under controlled thermodynamics conditions (*e. g.*, by varying temperature and pressure values). We have systematically analyzed how the transport dynamics of the confined nanofluids depend on the interlayer distances and the role of the oxide functional groups. Our results show the water flux is much more effective for graphene than for graphene oxide membranes. These results are attributed to the H-bonds formation between oxide functional groups and water, which traps the water molecules and precludes ultrafast water transport through the nanochannels.

## INTRODUCTION

Recent studies have revealed ultra-fast water transport in graphene-derived functional membranes in addition to their high selectivity [1] In particular, graphene oxide (GO), composed of graphene layers decorated with oxygen based functional groups, appears to be an excellent candidate for fluid filtration and gas separation applications. The fast water transport in carbon materials opens the perspective to develop membranes with both high selectivity and high flux values. Geim *et al*. reported that GO could be completely impermeable to liquids, vapours and gases, yet allowing unconstrained water permeation [2,3]. Although these selective mechanisms remain unclear, the fast permeation through GO membranes was attributed to ultralow friction monolayer water flow through two-dimensional pristine graphene channels [4]. The GO oxidized regions are considered to act as spacers and a minimum interlayer distance seems to be necessary for the capillary mechanism to work [2-4]. A capillary driven force was also proposed to explain these phenomena. These findings have led to large number of theoretical papers trying to address the mechanisms behind the observed fast water permeation. Conflicting results were reported on the origin of water flow mechanisms and it has been proposed that hydrogen bonds between water molecules and functional groups of GO sheets play a fundamental role [5-7] to determine the rate permeation values.

These membranes exhibit very complex topologies and the used experimental techniques to fabricate them can lead to distinct micro and nano morphologies and transport pathways [2-4]. In this work, we propose structural models to mimic membranes composed of different multiple layered graphene-based (pristine graphene and graphene oxide) sheets into contact with a water reservoir. We carried out a detailed investigation of the dynamics of water permeation through these model membranes using fully atomistic molecular dynamics simulations.

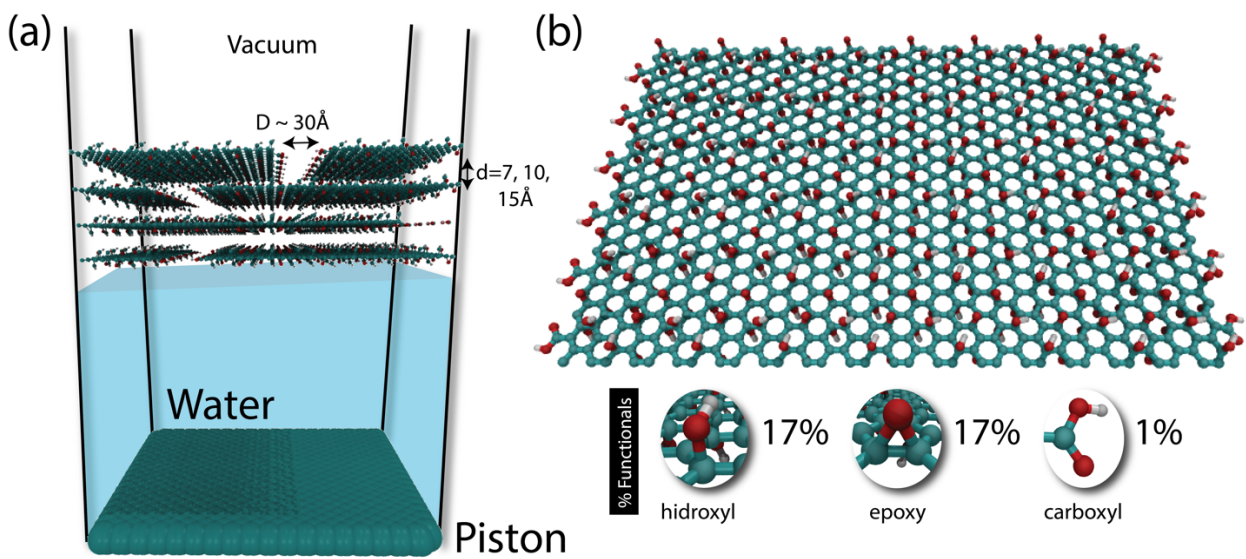

**Figure 1**. (a) The simulated system composed of multilayer graphene-based sheets into contact with a water reservoir at constant temperature and pressure values, which are controlled by a movable piston on the bottom. (b) Typical used graphene oxide sheet composed of: 17% of hydroxyl, 17% of epoxy and 1% of carboxyl groups.

**THEORY**

Full atomistic classical molecular dynamics (MD) simulations were carried out to study the water permeation into multilayer graphene-based membranes. In Figure 1a we present the schematic simulation set up. Typical structural models contain four graphene sheets perforated with nanoslits of width $D \sim 30$ Å and parallel arranged by a $d$ distance. Three d different values were considered here: 7, 10 and 15 Å. The simulation box dimensions are 93.6 x 44.2 x 200 Å$^3$. Two types of graphene sheets were considered: pristine graphene (pG) and graphene oxide (GO) with 35% content of oxygen (from hydroxyl, epoxy and carboxyl groups) atoms, as shown in Figure 1b. To build GO sheet, the graphene was functionalized with hydroxyl and epoxy functional groups on the both sides and with the carboxyl functional on the sheet edges.

The system configuration was then prepared by placing a water liquid reservoir into contact to the fixed membrane. The reservoir contains 8190 water molecules and mass density ~1 g/cm$^3$. The initial water configuration was generated using the Packmol [8] code and equilibrated at ambient pressure and temperature through molecular dynamics simulations. Once the reservoir is placed into contact to the membrane, the water flow was simulated through controlling the reservoir thermodynamics properties. The reservoir temperature is kept constant and equals to 300 K using a Nosé-Hoover thermostat [9,10] and the pressure is controlled by a movable piston of graphene placed on the bottom of the simulation box. The piston allows the reservoir volume to vary while the water moves into the membrane. The piston position is scaled by the force experienced by the piston as consequence of the reaction water force. This protocol is very effective to mimic an infinite water reservoir.

We used CHARMM [11,12] Lennard-Jones force field parameters for both pG and GO. Partial charge values were taken from Ref. [13]. The rigid extended simple point charge (SPC/E) model [14] was used to describe water molecules. Periodic boundary conditions were imposed along the *xy*-plane containing the membranes, while fixed boundaries were kept along the *z*-direction. The molecular dynamics simulations were carried out using the open source software called large-scale parallel molecular dynamics simulation code (LAMMPS) [15].

**DISCUSSION**

The water permeation through graphene-based membrane was investigated by performing MD simulations of graphene-based multilayers membranes (hereafter just called membranes) into contact with liquid water reservoir at constant pressure P (which can vary), temperature T = 300 K, and mass density ~1 g/cm$^3$. Figure 2a shows the time evolution of the normalized number of water molecules flowing out of the membrane for six different configurations: pG membranes with different interlayer distances: 7 Å (pG7), 10 Å (pG10), and 15 Å (pG15) and GO membranes for the same 3 interlayer distances referred as GO7, GO10, G15. The water permeation through graphene-based membranes is driven by a difference of pressure between the inlet of the membrane (at the bottom), which is under the water reservoir pressure and the outlet of the membrane (on top), which is vacuum. The difference of pressure is the same for each considered case. A water flow upwards is then induced, as illustrated in Figure 2 (right side), which shows some MD snapshots of graphene and GO membranes at three different MD times: 100, 500 and 2000 ps (I, II, and III).

Figure 2a shows two different regimes: the first one suggests a continuous flux of water crossing the membrane with the water flux varying linearly with time. Therefore, the flux (number of molecules per picosecond) can be computed by fitting a linear equation (y=ax+b), where *a* represents the water flux and –b/a is related to the time required for the water molecules to completely wet the membrane. Table 1 displays the water flux for each case considered. We can infer two important conclusions from these results: *i)* the flow rate (parameter "*a*") increases with interlayer distances; *ii)* the water flows inside pG is much faster than inside GO. The first observation suggests that, at this dimension scale, the spatial constrain effects are much more important than the capillarity effects. Indeed, at very narrow channels, such as in the case of d=7 Å, the water flux is reduced due to hysteric hindrance effects. In this configuration, the molecules are restricted to form only a single water layer, which forces the water diffusion to be essentially two-dimensional. This arrangement is depicted in the inset of Figure 3a. Further plateau regimes in Figure 2a correspond to the cases when the reservoir becomes completely empty; see MD snapshots presented in right side of Figure 2. This regime results from our adopted model, which tries to mimic an infinity water reservoir through a finite size one. We carried out some test to confirm that the finite size model does not affect our main conclusions.

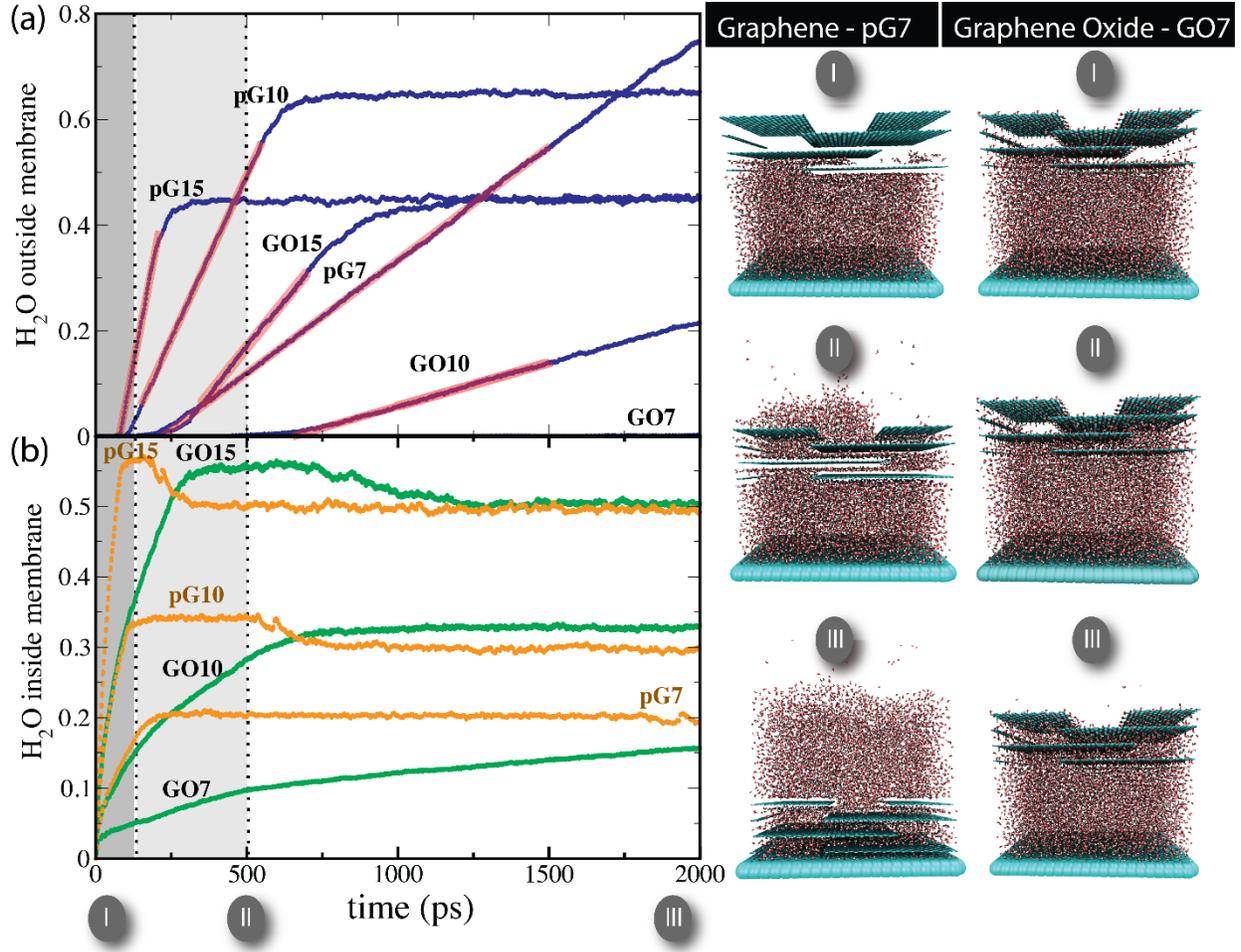

**Figure 2**. (a) Normalized number of water molecules flowing out of the membrane and; (b) the normalized number of water molecules inside the membrane as function of time. The number of water molecules inside each pG and GO membrane is the same. The snapshots of simulation box of pG7 and GO7 at three different moments, 100, 500 and 2000 ps, are displayed in the Figure right side.

**Table 1**. Water flux for the different systems (see text for discussions)

|  | GO7 | GO10 | GO15 | Graphene7 | Graphene10 | Graphene15 |
|---|---|---|---|---|---|---|
| Flux (#mol/ps) | 0.016 | 1.35 | 5.78 | 3.5 | 10.2 | 24.2 |

Comparing the water flux between pG and GO membranes (Figure 2), the water permeates inside graphene membrane much faster than inside GO, which is associated to the frictionless feature of graphene [2]. Graphene is hydrophobic; thus, in principle, this would prevent water to get inside of the membrane. However, once the water is inside, it strongly repeals the water. Due to the pressure difference of, *i.e.*, the reservoir pressure is much higher than the outside vacuum; the water can cross the membrane with an extremely high flux rate. In contrast, GO is more hydrophilic. In spite of the good affinity of GO to water, our simulations show that the water wet the membrane much slower than in graphene (see Figure 2b). Indeed, this strong interaction is responsible to form water agglomerates that block the entrance of the channels. This goes in agreement with theoretical results presented in the literature [7] and in contrast to the idea of attributing GO as "miracle material" with very fast water permeation.

To understand the water dynamics mechanism inside the membrane, a careful inspection on the organization (how the molecules are spatially arranged) of water molecules inside the membrane was performed. In this analysis, the system is kept in equilibrium (T=300 K) with no water flux. Figure 3a shows the probability distribution $P(cos\theta)$, where $\theta$ is the angle between the membrane normal vector ($\hat{z}$) and the normal vector to the water molecular plane ($\hat{n}$). Our results show that the probability to find coplanar water molecules is much higher inside graphene membrane in comparison to GO. The apparent water disorder inside GO can be attributed to the strong H-bonds formed between water molecules and the functional groups presented in GO, as shown in the MD snapshot of Figure 3b. As argued before, the water can easily slip inside graphene (hydrophobic), while the GO (hydrophilic) traps the water molecules. The inset of Figure 3a shows the structure of monolayer water inside GO7 and pG7.

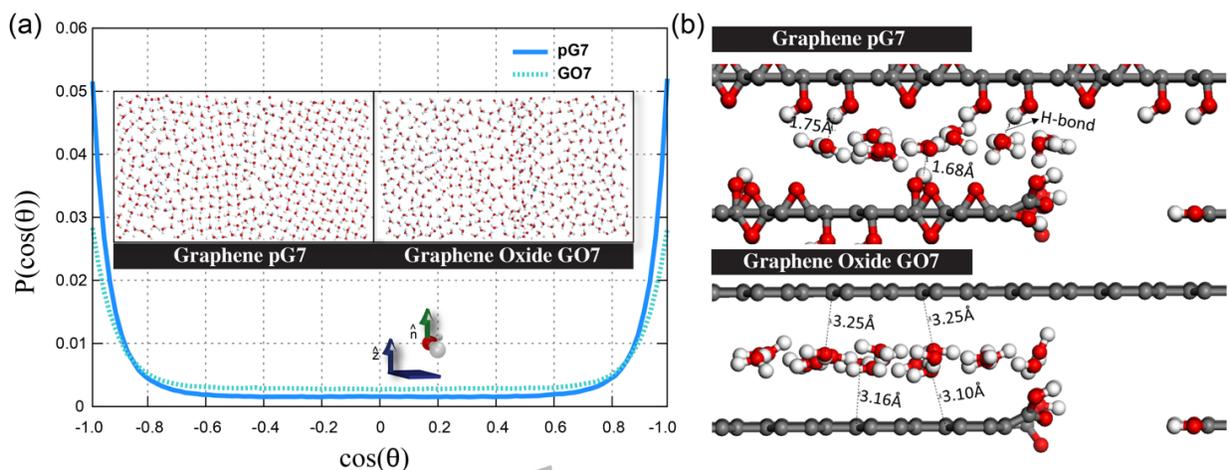

**Figure 3.** (a) Probability function distribution of cos(θ), where θ corresponds to the angles between the membrane surface normal and the normal vector to the water molecular plane. (b) MD snapshot showing the H-bonds formed between water and GO functional groups.

**CONCLUSIONS**

Through a fully atomistic classical molecular dynamics (MD) simulations we have investigated the dynamics of water permeation in pristine graphene and GO membranes. The simulated systems are composed of multiple layered graphene-based sheets into contact with a water reservoir under constant pressure and temperature = 300 K. In this work, we have systematically analyzed how the transport dynamics of the confined water depends on the interlayer distances. Our findings show the water flux through pristine graphene membranes are much higher in comparison to GO membranes, for all interlayer distances investigated here: $d$=7, 10, 15 Å. These results are attributed to the H-bonds formation between oxide functional groups and water, which traps the water molecules and prevents ultrafast water transport through the channels. Our results are consistent with the available experimental date and contribute to clarify some important aspects of the confined water behavior in GO membranes.

**ACKNOWLEDGMENTS**

This work was supported in part by the Brazilian Agencies CAPES, CNPq and FAPESP. The authors also thank the Center for Computational Engineering and Sciences at Unicamp for financial support through the FAPESP/CEPID Grant # 2013/08293-7. CFW thanks São Paulo Research Foundation (FAPESP) Grant # 2016/12340-9 for financial support.

**REFERENCES**
1. G. Liu, W. Jin, and N. Xu, *Chem. Soc. Rev.* **44**, 5016 (2015).
2. R. R. Nair, H. A. Wu1, P. N. Jayaram, I. V. Grigorieva, and A. K. Geim., *Science* **335**, 442 (2012).
3. R. K. Joshi, P. Carbone, F. C. Wang, V. G. Kravets, Y. Su, I. V. Grigorieva, H. A. Wu, A. K. Geim, and R. R. Nair, *Science* **343**, 752 (2014).
4. K. Falk, F. Sedlmeier, L. Joly, R. R. Netz, and L. Bocquet, *Nano Lett.* **10**, 4067 (2010).
5. N. Wei, X. Peng, and Z. Xu, *ACS Appl. Mat Int.* **6**, 5877 (2014).
6. N. Wei, X. Peng, and Z. Xu, *Phys. Rev. E* **89**, 012113 (2014).
7. J. A. L. Willcox and H. J. Kim, *ACS Nano* (2017), DOI: 10.1021/acsnano.6b08538.
8. L. Martínez, *J. Comp. Chem.* **30**, 2157 (2009).
9. S. Nosé, *J. Chem. Phys.* **81**, 511 (1984).
10. W. G. Hoover, *Phys. Rev. A* **31**, 1695 (1985).
11. K. Vanommeslaeghe, E. Hatcher, C. Acharya, S. Kundu, S. Zhong, J. Shim, E. Darian, O. Guvench, P. Lopes, I. Vorobyov, A. D. MacKerell Jr., *J. Comp. Chem.* **31**, 671 (2010).
12. W. Yu, X. He, K. Vanommeslaeghe, and A. D. MacKerell Jr., *J. Comp. Chem.*, **33**, 2451 (2012).
13. S. Jiao, and Z. Xu, *ACS Appl. Mater. Interfaces*, **7**, 9052 (2015).
14. H. J. C. Berendsen, J. R. Grigera, T. P. Straatsma, *J. Phys. Chem.* **91**, 6269 (1987).
15. S. Plimpton, *J. Comput. Phys.* **117**, 1 (1995).